\documentstyle[12pt,twoside]{article}

\def\a{\alpha}
\def\b{\beta}

\def\d{\delta}
\def\e{\epsilon}                
\def\f{\phi}                    
\def\g{\gamma}
\def\h{\eta}

\def\j{\psi}

\def\l{\lambda}
\def\m{\mu}
\def\n{\nu}

\def\p{\pi}                     
\def\s{\sigma}                  

\def\x{\xi}

\def\F{\Phi}
\def\G{\Gamma}

\def\L{\Lambda}

\def\S{\Sigma}

\def\cl{{\cal L}}

\catcode`@=11

\def\un#1{\relax\ifmmode\@@underline#1\else $\@@underline{\hbox{#1}}$\relax\fi}

{\catcode`\'=\active \gdef'{{}~\bgroup\prim@s}}

\def\magstep#1{\ifcase#1 \@m\or 1200\or 1440\or 1728\or 2074\or 2488\or
        2986\fi\relax}

\font\twfvmi=cmmi10\@magscale5
    \skewchar\twfvmi='177
\font\twfvsy=cmsy10\@magscale5
    \skewchar\twfvsy='60
\font\twfvly=lasy10\@magscale5
\font\thtyrm=cmr10\@magscale6

\def\vpt{\textfont\z@\fivrm
  \scriptfont\z@\fivrm \scriptscriptfont\z@\fivrm
\textfont\@ne\fivmi \scriptfont\@ne\fivmi \scriptscriptfont\@ne\fivmi
\textfont\tw@\fivsy \scriptfont\tw@\fivsy \scriptscriptfont\tw@\fivsy
\textfont\thr@@\tenex \scriptfont\thr@@\tenex \scriptscriptfont\thr@@\tenex
\def\prm{\fam\z@\fivrm}%
\def\unboldmath{\everymath{}\everydisplay{}\@nomath
  \unboldmath\fam\@ne\@boldfalse}\@boldfalse
\def\boldmath{\@subfont\boldmath\unboldmath}%
\def\pit{\@getfont\pit\itfam\@vpt{cmti5}}%
\def\psl{\@subfont\sl\it}%
\def\pbf{\@getfont\pbf\bffam\@vpt{cmbx5}}%
\def\ptt{\@subfont\tt\rm}%
\def\psf{\@subfont\sf\rm}%
\def\psc{\@subfont\sc\rm}%
\def\ly{\fam\lyfam\fivly}\textfont\lyfam\fivly
    \scriptfont\lyfam\fivly \scriptscriptfont\lyfam\fivly
\@setstrut\rm}

\def\@vpt{}

\def\vipt{\textfont\z@\sixrm
  \scriptfont\z@\sixrm \scriptscriptfont\z@\sixrm
\textfont\@ne\sixmi \scriptfont\@ne\sixmi \scriptscriptfont\@ne\sixmi
\textfont\tw@\sixsy \scriptfont\tw@\sixsy \scriptscriptfont\tw@\sixsy
\textfont\thr@@\tenex \scriptfont\thr@@\tenex \scriptscriptfont\thr@@\tenex
\def\prm{\fam\z@\sixrm}%
\def\unboldmath{\everymath{}\everydisplay{}\@nomath
  \unboldmath\@boldfalse}\@boldfalse
\def\boldmath{\@subfont\boldmath\unboldmath}%
\def\pit{\@subfont\it\rm}%
\def\psl{\@subfont\sl\rm}%
\def\pbf{\@getfont\pbf\bffam\@vipt{cmbx6}}%
\def\ptt{\@subfont\tt\rm}%
\def\psf{\@subfont\sf\rm}%
\def\psc{\@subfont\sc\rm}%
\def\ly{\fam\lyfam\sixly}\textfont\lyfam\sixly
    \scriptfont\lyfam\sixly \scriptscriptfont\lyfam\sixly
\@setstrut\rm}

\def\@vipt{}

\def\xxxpt{\textfont\z@\thtyrm
  \scriptfont\z@\twfvrm \scriptscriptfont\z@\twtyrm
\textfont\@ne\twfvmi \scriptfont\@ne\twfvmi \scriptscriptfont\@ne\twtymi
\textfont\tw@\twfvsy \scriptfont\tw@\twfvsy \scriptscriptfont\tw@\twtysy
\textfont\thr@@\tenex \scriptfont\thr@@\tenex \scriptscriptfont\thr@@\tenex
\def\unboldmath{\everymath{}\everydisplay{}\@nomath\unboldmath
        \textfont\@ne\twfvmi \textfont\tw@\twfvsy \textfont\lyfam\twfvly
        \@boldfalse}\@boldfalse
\def\boldmath{\@subfont\boldmath\unboldmath}%
\def\prm{\fam\z@\thtyrm}%
\def\pit{\@subfont\it\rm}%
\def\psl{\@subfont\sl\rm}%
\def\pbf{\@getfont\pbf\bffam\@xxxpt{cmbx10\@magscale6}}%
\def\ptt{\@subfont\tt\rm}%
\def\psf{\@subfont\sf\rm}%
\def\psc{\@subfont\sc\rm}%
\def\ly{\fam\lyfam\twfvly}\textfont\lyfam\twfvly
   \scriptfont\lyfam\twfvly \scriptscriptfont\lyfam\twtyly
\@setstrut \rm}

\def\@xxxpt{}

\def\Huge{\@setsize\Huge{36pt}\xxxpt\@xxxpt}

\font\thtymi=cmmi10\@magscale6
    \skewchar\thtymi='177
\font\thtysy=cmsy10\@magscale6
    \skewchar\thtysy='60
\font\thtyly=lasy10\@magscale6
\font\thsirm=cmr12\@magscale6

\def\xxxvipt{\textfont\z@\thsirm
  \scriptfont\z@\thtyrm \scriptscriptfont\z@\twfvrm
\textfont\@ne\thtymi \scriptfont\@ne\thtymi \scriptscriptfont\@ne\twfvmi
\textfont\tw@\thtysy \scriptfont\tw@\thtysy \scriptscriptfont\tw@\twfvsy
\textfont\thr@@\tenex \scriptfont\thr@@\tenex \scriptscriptfont\thr@@\tenex
\def\unboldmath{\everymath{}\everydisplay{}\@nomath\unboldmath
        \textfont\@ne\thtymi \textfont\tw@\thtysy \textfont\lyfam\thtyly
        \@boldfalse}\@boldfalse
\def\boldmath{\@subfont\boldmath\unboldmath}%
\def\prm{\fam\z@\thsirm}%
\def\pit{\@subfont\it\rm}%
\def\psl{\@subfont\sl\rm}%
\def\pbf{\@getfont\pbf\bffam\@xxxpt{cmss12\@magscale6}}%
\def\ptt{\@subfont\tt\rm}%
\def\psf{\@subfont\sf\rm}%
\def\psc{\@subfont\sc\rm}%
\def\ly{\fam\lyfam\thtyly}\textfont\lyfam\thtyly
   \scriptfont\lyfam\thtyly \scriptscriptfont\lyfam\twfvly
\@setstrut \rm}

\def\@xxxvipt{}

\def\HUGE{\@setsize\HUGE{43pt}\xxxvipt\@xxxvipt}

\catcode`@=12

\font\tenex=cmex10 scaled 1200

\def\Sc#1{\hbox{\sc #1}}        
\font\oo=circlew10            

\def\bo{{\raise.05ex\hbox{\large$\Box$}\:}}             
\def\cbo{{\,\raise-.15ex\Sc [\,}}                       
\def\pa{\partial}                                       
\def\su{\sum}                                           
\def\TH{{\raise.2ex\hbox{$\displaystyle \bigodot$}\mskip-4.7mu \llap H \;}}
\def\face{\hbox{\normalsize$\;\;\:{\raise.9ex\hbox{\oo n}\mskip-13mu \llap
        {${\buildrel{\hbox{\frtnrm ..}}\over\smile}$}}\:$}}     
\def\Face{{\raise.2ex\hbox{$\displaystyle \bigodot$}\mskip-2.2mu \llap {$\ddot
        \smile$}}}                                      
\def\Lhat{{\bf\rlap{\kern-.09em$\hat{\phantom L}$}L}}
\def\Lcheck{{\bf\rlap{\kern-.09em$\check{\phantom L}$}L}}

\def\sp#1{{}^{#1}}                              
\def\sb#1{{}_{#1}}                              
\def\sl#1{\rlap{\hbox{$\mskip 1 mu /$}}#1}      
\def\sket#1{\left| #1\right\rangle}             
\def\leftrightarrowfill{$\mathsurround=0pt \mathord\leftarrow \mkern-6mu
        \cleaders\hbox{$\mkern-2mu \mathord- \mkern-2mu$}\hfill
        \mkern-6mu \mathord\rightarrow$}
\def\dvec#1{\vbox{\ialign{##\crcr
        \leftrightarrowfill\crcr\noalign{\kern-1pt\nointerlineskip}
        $\hfil\displaystyle{#1}\hfil$\crcr}}}           
\def\dt#1{{\buildrel {\hbox{\LARGE .}} \over {#1}}}     
\def\ddt#1{{\buildrel {\hbox{\LARGE .\kern-2pt.}} \over {#1}}}

\def\frac#1#2{{\textstyle{#1\over\vphantom2\smash{\raise.20ex
        \hbox{$\scriptstyle{#2}$}}}}}                   
\def\ha{\frac12}                                        
\def\sfrac#1#2{{\vphantom1\smash{\lower.5ex\hbox{\small$#1$}}\over
        \vphantom1\smash{\raise.4ex\hbox{\small$#2$}}}} 
\def\bfrac#1#2{{\vphantom1\smash{\lower.5ex\hbox{$#1$}}\over
        \vphantom1\smash{\raise.3ex\hbox{$#2$}}}}       
\def\afrac#1#2{{\vphantom1\smash{\lower.5ex\hbox{$#1$}}\over#2}}    

\def\boxes#1{
        \newcount\num
        \num=1
        \newdimen\downsy
        \downsy=-1.64ex
        \mskip-7.8mu
        \bo
        \loop
        \ifnum\num<#1
        \llap{\raise\num\downsy\hbox{$\bo$}}
        \advance\num by1
        \repeat}
\def\boxup#1#2{\newcount\numup
        \numup=#1
        \advance\numup by-1
        \newdimen\upsy
        \upsy=.82ex
        \mskip7.8mu
        \raise\numup\upsy\hbox{$#2$}}

\newskip\humongous \humongous=0pt plus 1000pt minus 1000pt
\def\caja{\mathsurround=0pt}

\newif\ifdtup
\def\panorama{\global\dtuptrue \openup2\jot \caja
        \everycr{\noalign{\ifdtup \global\dtupfalse
        \vskip-\lineskiplimit \vskip\normallineskiplimit
        \else \penalty\interdisplaylinepenalty \fi}}}
\def\li#1{\panorama \tabskip=\humongous                         
        \halign to\displaywidth{\hfil$\displaystyle{##}$
        \tabskip=0pt&$\displaystyle{{}##}$\hfil
        \tabskip=\humongous&\llap{$##$}\tabskip=0pt
        \crcr#1\crcr}}

\def\NP{Nucl. Phys. B}
\def\PL{Phys. Lett. }

\def\PRD{Phys. Rev. D}
\def\ref#1{$\sp{#1]}$}

\parskip=\medskipamount                 
\def\baselinestretch{1.2}       
\thicklines                         
\def\title#1#2#3#4{
\begin{document}
        {\hbox to\hsize{#4 \hfill QMW/PH/ #3}}\par
        \begin{center}\vskip.5in minus.1in {\Large\bf #1}\\[.5in minus.2in]{#2}
        \vskip1.4in minus1.2in {\bf ABSTRACT}\\[.1in]\end{center}
        \begin{quotation}\par}
\def\author#1#2{#1\\[.1in]{\it #2}\\[.1in]}
\def\AM{Aleksandar Mikovi\'c\,\footnote
   {Work supported by the U.K. Science and Engineering Research Council}
\footnote{E-mail address: MIKOVIC@V1.PH.QMW.AC.UK}
\\[.1in] {\it Department of Physics, Queen Mary and Westfield
College,\\ Mile End Road, London E1 4NS, U.K.}\\[.1in]}
\def\WS{W. Siegel\\[.1in] {\it Institute for Theoretical
        Physics,\\ State University of New York, Stony Brook, NY 11794-3840}
        \\[.1in]}
\def\endtitle{\par\end{quotation}\vskip3.5in minus2.3in\newpage}


\def\endabstract{\par\end{quotation}
        \renewcommand{\baselinestretch}{1.2}\small\normalsize}


\def\xpar{\par}                                         
\def\letterhead{
        \centerline{\large\sf QUEEN MARY AND WESTFIELD COLLEGE}
        \centerline{\sf Department of Physics}
        \vskip-.07in
        \centerline{\sf Mile End Road, London E1 4NS}
        \rightline{\scriptsize\sf Dr. Aleksandar Mikovi\'c}
        \vskip-.07in
        \rightline{\scriptsize\sf Tel: 071-975-5055}
        \vskip-.07in
        \rightline{\scriptsize\sf E-mail: MIKOVIC@V1.PH.QMW.AC.UK}}
\def\sig#1{{\leftskip=3.75in\parindent=0in\goodbreak\bigskip{Sincerely yours,}
\nobreak\vskip .7in{#1}\par}}


\def\ree#1#2#3{
        \hfuzz=35pt\hsize=5.5in\textwidth=5.5in
        \begin{document}
        \ttraggedright
        \par
        \noindent Referee report on Manuscript \##1\\
        Title: #2\\
        Authors: #3}


\def\start#1{\pagestyle{myheadings}\begin{document}\thispagestyle{myheadings}
        \setcounter{page}{#1}}


\catcode`@=11

\def\ps@myheadings{\def\@oddhead{\hbox{}\footnotesize\bf\rightmark \hfil
        \thepage}\def\@oddfoot{}\def\@evenhead{\footnotesize\bf
        \thepage\hfil\leftmark\hbox{}}\def\@evenfoot{}
        \def\sectionmark##1{}\def\subsectionmark##1{}
        \topmargin=-.35in\headheight=.17in\headsep=.35in}
\def\ps@acidheadings{\def\@oddhead{\hbox{}\rightmark\hbox{}}
        \def\@oddfoot{\rm\hfil\thepage\hfil}
        \def\@evenhead{\hbox{}\leftmark\hbox{}}\let\@evenfoot\@oddfoot
        \def\sectionmark##1{}\def\subsectionmark##1{}
        \topmargin=-.35in\headheight=.17in\headsep=.35in}

\catcode`@=12

\def\sect#1{\bigskip\medskip\goodbreak\noindent{\large\bf{#1}}\par\nobreak
        \medskip\markright{#1}}
\def\chsc#1#2{\phantom m\vskip.5in\noindent{\LARGE\bf{#1}}\par\vskip.75in
        \noindent{\large\bf{#2}}\par\medskip\markboth{#1}{#2}}
\def\Chsc#1#2#3#4{\phantom m\vskip.5in\noindent\halign{\LARGE\bf##&
        \LARGE\bf##\hfil\cr{#1}&{#2}\cr\noalign{\vskip8pt}&{#3}\cr}\par\vskip
        .75in\noindent{\large\bf{#4}}\par\medskip\markboth{{#1}{#2}{#3}}{#4}}
\def\chap#1{\phantom m\vskip.5in\noindent{\LARGE\bf{#1}}\par\vskip.75in
        \markboth{#1}{#1}}
\def\refs{\bigskip\medskip\goodbreak\noindent{\large\bf{REFERENCES}}\par
        \nobreak\bigskip\markboth{REFERENCES}{REFERENCES}
        \frenchspacing \parskip=0pt \renewcommand{\baselinestretch}{1}\small}
\def\unrefs{\normalsize \nonfrenchspacing \parskip=medskipamount}
\def\Item{\par\hang\textindent}
\def\Itemitem{\par\indent \hangindent2\parindent \textindent}
\def\makelabel#1{\hfil #1}
\def\topic{\par\noindent \hangafter1 \hangindent20pt}
\def\Topic{\par\noindent \hangafter1 \hangindent60pt}


\title{Exactly Solvable Models of 2d Dilaton Quantum Gravity}
{\AM}{92/12}{June 1992}
We study canonical quantization of a class of 2d dilaton gravity models,
which
contains the model proposed by Callan, Giddings, Harvey and Strominger.
A set of non-canonical phase space
variables is found, forming an $SL(2,{\bf R}) \times U(1)$
current algebra, such that the
constraints become quadratic in these new variables. In the case when the
spatial manifold is compact, the corresponding quantum theory can be
solved exactly, since it reduces to a problem of
finding the cohomology of a free-field Virasoro algebra. In the non-compact
case, which is relevant for 2d black holes,
this construction is likely to break down,
since the most general field configuration
cannot be expanded into Fourier modes. Strategy for circumventing this problem
is discussed.

\endtitle

\sect{1. Introduction}

Recently there has been a lot of interest in two-dimensional renormalisible
models of gravity coupled to scalar fields. These are relevant for
non-critical string theory \cite{{kpz},{ddk}},
as well as toy models for describing the formation
and evaporation of black holes \cite{cghs}. As shown in \cite{tsey}, the
most general form of the action for such a model is
$$ S= -\int_{M} d^2 x \sqrt{-g}( \ha g\sp{\m\n}\pa\sb \m\f\pa\sb \n \f +
\ha\a R\f + V(\f)) \quad,\eqno(1.1)$$
where $M$ is a 2d manifold,
$g\sb{\m\n}$ is a metric on $M$, $\f$ is a scalar field (dilaton),
$\a$ is a constant (background charge) and $R$ is the 2d curvature scalar.
We will label the time coordinate $x^0 = t$ and the space coordinate
$x^1 = x$, while the corresponding derivatives will be denoted as $\dt{}$
and $'$, respectively.

The form (1.1) can be always achieved after suitable field redefinitions
\cite{tsey}. For example, the CGHS action
$$ S= -\frac18 \int_{M} d^2 x \sqrt{-g}e^{-2\F}( R +
4 g\sp{\m\n}\pa\sb \m\F\pa\sb \n \F + c ) \quad,\eqno(1.2)$$
takes the form (1.1), with $V(\f)={c\over 8}e^{\f/\a}$,
after the following field redefinitions
$$ \f = {1\over 4\a} e^{-2\F} \quad,\quad g_{\m\n} \to {1\over4\a\f}e^{\f/\a}
g_{\m\n} \quad.\eqno(1.3)$$
Depending on the
form of the potential $V(\f)$ and whether the spatial section of the
2d manifold $M$ is compact or
non-compact, one can get models describing a non-critical string theory or 2d
black holes. One loop perturbative analysis of (1.1) has been carried out in
\cite{tsey}, where it was pointed out that the analysis symplify in the
case when $V(\f)=\L e^{\b \f}$, where $\L$ and $\b$ are constants.

Canonical quantization methods are more successful in exploring
the nonperturbative nature of
quantum gravity in four space-time dimensions \cite{asht}
then the standard path-integral methods. This may naturaly lead one to
apply the same methods in the case of 2d gravity, more specifically, to
the theory defined by the action (1.1). The canonical analysis in the
case $\b=0$ and compact spatial manifold
has been already carried out by the author \cite{mik}, where
it was demonstrated that the corresponding quantum theory is exactly
solvable. This was achieved by using non-canonical phase space variables,
forming an $SL(2, {\bf R})\times U(1)$ Kac-Moody algebra,
which transformed the constraints
into quadratic polynomials of the new variables. By using the free-field
realization of the $SL(2,{\bf R})$ currents, the constraints became
a free-field
realization of the Virasoro algebra, whose cohomology is known \cite{bmp}.

In this paper we show that the same can be done in the case $\b \ne 0$
and the spatial manifold is compact.
Adding conformaly coupled matter does not change this result, which means that
the physical Hilbert space of the CGHS model on a circle
can be obtained by solving the cohomology of
a Virasoro algebra realised from $N+2$ free scalar fields,
with background charges, where N is the central charge of the conformally
coupled matter. Since one can very easily construct, level by level, the
physical Hilbert space for such systems (for $N=0$ there is a complete
solution \cite{bmp}), the model is exactly solvable.

Unfortunately, this construction breaks down in the non-compact case, due
to the absence of well defined Fouirer modes for a most general field
configuration. Namely, if one wants to include the black hole solutions
into the quantum theory, then one has to allow field configurations which
blow up either at $x=\pm\infty$ or at some finite $x$. Such
configurations cannot be expanded into Fouirer series. Therefore one needs
an alternative way of defining the quantum theory, which is discussed in
the conclussions.

\sect{2. Canonical Analysis}

The canonical formulation of (1.1) requires that the 2d manifold $M$ has a
topology of $\S \times {\bf R}$,
where $\S$ is the spatial manifold and ${\bf R}$ is the real line corresponding
to the time direction. $\S$ can be either a circle $S^1$ or a real line.
The compact spatial topology is relevant for string theory, while the
non-compact spatial topology is relevant for 2d black holes, although we will
argue at the end of the paper that the black hole solutions are possible even
in the compact case. The compact case is
simpler for analysis, due to absence of the ``surface" terms.
In the non-compact case, one can assume appropriate boundary conditions at
$x=\pm\infty$, such that boundary terms do not apear. However, in a most
general case they will be present.

Derivation of the canonical form of the action (1.1) is simplified by
introducing the laps function ${\cal N}(x,t)$ and the shift vector $n(x,t)$
\cite{asht}. Then the metric $g_{\m\n}$ takes the following form
$$g_{00}= -{\cal N}^2 + n^2 g \quad,\quad g_{01} = ng \quad,\quad g_{11}=g
\quad,\eqno(2.1)$$
where $g(x,t)$ is a metric on $\S$. After introducing the canonical momenta
for $g$ and $\f$ as
$$ p = {\pa \cl \over\pa \dt{g}} \quad,\quad
   \p = {\pa \cl \over\pa \dt{\f}} \quad,\eqno(2.2)$$
where $\cl$ is the Lagrangian density of (1.1), then up to surface terms,
the action becomes
$$ S= \int dt dx \left( p\dt{g} + \p\dt{\f} - {{\cal N}\over\sqrt{g}}G_0 -
n G\sb 1 \right) \quad, \eqno(2.3)$$
where
$$\li{G\sb 0 (x) &=  - {2\over{\a^2}}g^2 p^2
-{2\over\a}gp\p + \ha(\f^{\prime})^2 + gV(\f)
- {{\a}\over2}{g^{\prime}\over g}\f^{\prime} + \a\f^{\prime\prime}\cr
G\sb 1 (x) &= \p\f^{\prime} - 2p^{\prime}g - pg^{\prime}\quad.&(2.4)\cr}$$
The constraints $G_0$ and $G_1$ form a closed Poisson bracket algebra
$$\li{ \{ G_0 (x), G_0 (y) \} &= -\d^{\prime} (x-y)(G_1 (x) + G_1 (y)) \cr
\{ G_1 (x), G_0 (y) \} &= -\d^{\prime} (x-y)(G_0 (x) + G_0 (y)) \cr
\{ G_1 (x), G_1 (y) \} &= -\d^{\prime} (x-y)(G_1 (x) + G_1 (y)) \quad,&(2.5)
\cr}$$
where the fundamental Poisson brackets are defined as
$$\{p(x),g(y)\}= \d (x -y)
\quad,\quad
\{\p(x),\f(y)\} = \d (x -y)\quad.\eqno(2.6)$$
$G_1$ generates the spatial diffeomorphisms, while $G_0$
generates the time translations of $\S$, in full analogy with the $3+1$
gravity case.
Note that the algebra (2.5) is isomorphic to two comuting copies of the
1d diffeomorphism algebra, which can be seen by defining the constraints as
$$ T_{\pm} = \ha ( G_0 \pm G_1 ) \quad.\eqno(2.7)$$

Introduction of the conformally coupled scalar matter changes the action (1.1)
by
$$S_m = -\ha\int_{M} d^2 x \sqrt{-g} g\sp{\m\n}\pa\sb \m\f^i\pa\sb \n \f_i
\quad,\eqno(2.8)$$
where $i=1,...,N$. The constraints change as
$$\li{G_0 &\to  G_0 + \ha \p_i^2 + \ha (\f_i^{\prime})^2\cr
G\sb 1  &\to G_1 + \p^i\f^{\prime}_i \quad,&(2.9)\cr}$$
where $\p_i$ are the canonically conjugate momenta for $\f_i$.

Since we are dealing with a reparametrization invariant system, the
Hamiltonian vanishes on the constraint surface (i.e. it is proportional
to the constraints). Therefore the dynamics is determined by
the constraints only. Since $G_0$ and $G_1$ are irreducible, there will be
$(2+N) - 2 = N$ local physical degrees of freedom. When $N=0$, there are
only finitely many global physical degrees of freedom (zero modes of $g$ and
$\f$),
and one is dealing with
a topological field theory. When $N\ne 0$, these global degrees of freedom
will be present, together with the local ones. In the quantum theory, this
classical counting can be spoiled by the anomalies. However, when the anomalies
are absent, this counting should still hold, as the subsequent analysis will
show.

\sect{3. $SL(2,{\bf R})\otimes U(1)$ Variables}

We now specialize to the case $V(\f)=\L e^{\b\f}$.
As in the case $\b=0$ \cite{mik}, the variables $(g,p,\f,\p)$ are not
convinient for quantization, since $G_0$ is a non-polynomial function of
these variables. First we perform a canonical transformation in order to
get rid off the exponential in $\f$ term
$$ \li{ g = e^{-\b\tilde\f}\tilde g \quad,&\quad p = e^{\b\tilde\f} \tilde p
\cr \f = \tilde \f \quad,&\quad \p = {\tilde \p} + \b \tilde p \tilde g
\quad.&(3.1)\cr}$$
The constraints now become
$$\li{G\sb 0 (x) &=  - {2\over{\a^2}}(1+\a\b)g^2 p^2
-{2\over\a}gp\p + \ha(1+\a\b)(\f^{\prime})^2 + \L g
- {{\a}\over2}{g^{\prime}\over g}\f^{\prime} + \a\f^{\prime\prime}\cr
G\sb 1 (x) &= \p\f^{\prime} - 2p^{\prime}g - pg^{\prime}\quad,&(3.2)\cr}$$
where we have dropped the tildas. As in the $\b=0$ case, we are going
to look for the analogs of the $SL(2,R)$ variables introduced in \cite{mik}.
We define
$$\li{(1+\a\b)J\sp + &= -{\sqrt{2}\over{g}} T\sb - + {\L\over\sqrt2}\cr
(1+\a\b)J\sp 0 &= (1+\a\b) gp + {\a\over2}\left( \p -
{\a\over2}{g^{\prime}\over g}\right)  \cr
J\sp - &= {\a^2\over\sqrt{2}} g \cr
(1+\a\b)^{\ha} P_D &= {1\over\sqrt{2}}\left( \p -
{\a\over 2} {g^{\prime}\over g} + (1 +\a\b)\f^{\prime} \right)
\quad. &(3.3)\cr}$$
The $(J^a ,P_D)$ variables satisfy an
$SL(2,{\bf R})\otimes U(1)$ current algebra
$$\li{ \{ J^a (x), J^b (y)\} &= f\sp{ab}\sb c J^c (x)
\d (x-y) - {\bar{\a}^2\over2} \h^{ab}\d^{\prime}(x-y) \cr
\{ P_D (x) , P_D (y) \} &= - \d^{\prime}(x-y)\quad,&(3.4)\cr}$$
where
$$ \bar{\a}^2 = {\a^2\over 1+\a\b }\quad,\eqno(3.5) $$
and $f\sp{ab}\sb c = 2\e\sp{abd}\h\sb{dc}$ with
$$\h^{ab} =\pmatrix{0 &0 &2\cr 0 &-1 &0 \cr 2 &0 &0\cr}\quad,\eqno(3.6)$$
and $\{J, P_D\} =0$. Instead of using the canonical variables $(\p_i,\f_i)$,
we introduce the left/right moving currents
$$P_i =  {1\over\sqrt{2}}\left( \p_i + \f_i^{\prime}\right) \quad,\quad
\tilde{P}_i =  {1\over\sqrt{2}}\left( \p_i - \f_i^{\prime}\right) \quad,
\eqno(3.7)$$
satisfying
$$\{ P_i (x) , P_j (y) \} = -\d_{ij} \d^{\prime}(x-y)\quad,\quad
\{ \tilde{P}_i (x) , \tilde{P}_j (y) \} = \d_{ij} \d^{\prime}(x-y)\quad,
\eqno(3.8)$$
and $\{ P, \tilde{P} \} = 0$.
Now one can show that the energy-momentum tensor
associated to the algebra (3.4) via the Sugavara construction, together
with the matter contribution
$$ {\cal S} = T_g + T_m $$
$$T\sb g = {1\over\bar{\a}^2}\h_{ab}J^a J^b - (J^0)^{\prime} \quad,
\quad T\sb m = \ha P_D^2 + {\bar{\a}\over\sqrt2}P_D^{\prime}
+ \ha P_i^2 \quad, \eqno(3.9)$$
satisfies $ {\cal S} = T_+ $ on the constraint surface.
Therefore the constraints become
$$\li{ J^+ (x) -\l &= 0 \cr
{\cal S}(x) = T\sb g (x) + T\sb m (x) &= 0\quad, &(3.10)\cr}$$
where $\l = {\L\over\sqrt2 (1+\a\b)}$.

\sect{4. Quantum Theory}

The quantum theory can be now constructed by following the approach of
\cite{mik}. We promote $J$'s and $P$'s
into Hermitian operators, satisfying
$$\li{ [ J^a(x), J^b(y)] &= if\sb{ab}\sp c J^c \d (x-y)
-i{k\over2}\h^{ab} \d^{\prime}(x-y ) \cr
 [ P_I (x) , P\sb J (y)] &= -i\d_{IJ}\d^{\prime} (x-y)
\quad,\quad I = i,D \quad.&(4.1)\cr}$$
We introduce a new constant $k$, which is different from $2\p\bar{\a}^2$
due to
ordering ambiguities. It will be determined from the requiriment of anomally
cancelation. Since the constraints are independent of the $\tilde{P}_i$
variables, then the complete physical Hilbert space will be a tensor product
of a $\tilde{P}$ Hilbert space with the physical Hilbert space of $(J, P_I)$
variables.

In the case of $\S=S^1$, we construct the kinematical Hilbert space as a
Fock space
built on the vacuum state anhilliated by the positive Fouirer modes of $J$ and
$P_I$. If we define the Fourier modes as
$$J^a (x) = {1\over 2\p}\su_n e^{i n x}J^a_n \quad,\quad
P_I (x) = {1\over\sqrt{2\p}}\su_n e^{i n x}\a^I_n \quad,\eqno(4.2)$$
then (4.1) becomes
$$\li{ [ J^a_n, J^b_m] &= if\sb{ab}\sp c J^c_{n+m}
+{k\over2}\h^{ab}n \d_{n+m} \cr
 [ \a^I_n , \a^J_m ] &= \d_{IJ}n\d_{n+m}
\quad.&(4.3)\cr}$$
The Fock space vacuum is defined as $\sket{j,m}\otimes\sket{p_I}$, where
$\sket{j,m}$ is the $SL(2,{\bf R})$ vacuum
$$\li{J^a_n \sket{j,m} &= 0 \quad,\quad n\ge 1 \cr
  J^a_0 \sket{j,m} &= j^a \sket{j,m}\quad,&(4.4)\cr}$$
while $\sket{p_I}$ is the $U(1)$ vacuum
$$\li{\a_n^I \sket{p_M} &= 0 \quad,\quad n\ge 1 \cr
  \a_0^I \sket{p_M} &= p_I \sket{p_I}\quad.&(4.5)\cr}$$

The quantum constraints are defined as
$$L\sb n ={1\over k+2}\su_m \h_{ab} :J_{n-m}^a J_m^b : -in J_n^0
 +\ha \su_m :\a^I_{n-m}\a^I_m : + inQ_I \a^I_n \quad,\eqno(4.6)$$
where $S(x) ={1\over 2\p}\su_n e^{inx}L_n$, and the normal ordering is with
respect to the vacuum states (4.4-5). Note that the anomaly appears in the
quantum algebra of the constraints (4.6), proportional to the central charge
of gravity plus matter system
$$ c = {3k\over k + 2 } - 6k  + N + 1 + 12Q_D^2 \quad,\eqno(4.7)$$
where $Q_D = \sqrt{\p}\bar{\a}$. When anomally appears in the constraint
algebra, then one has to use the Gupta-Bleuler quantization procedure, or
the BRST quantization, which is more suitable in this case.

The BRST charge $\hat{Q}$ can be constructed as
$$ \hat{Q} = c\sb 0 ( L\sb 0 - a ) + \su_{n \ne 0} c\sb{-n} L\sb{n} +
\su_{n} c^+_{-n}(J^+_{n} - \l\d_{n,0}) + \cdots \quad,\eqno(4.8)$$
where $c_n$ and $c^+_n$ are the Fourier modes of the
ghosts corresponding to the constraints (3.10).
The dots correspond to the terms proportional to the ghost momenta, such that
$\hat{Q}^2 =0$, and $a$ is the intercept \cite{hik}. The
nilpotency condition requires vanishing of the total central charge,
which includes the ghost contributions
$$ c - 26 - 2 = 0 \quad,\eqno(4.9)$$
and the intercept must satisfy \cite{mik}
$$ a = 1 + {k\over4} - \ha Q_D^2 \quad.\eqno(4.10)$$

Evaluation of the cohomology of the BRST charge (4.8) simplifies if
one employs the
Wakimoto construction for the $SL(2,{\bf R})$ algebra (4.3) \cite{wak}.
As in the $\b=0$ case \cite{mik}, we
introduce three new variables $\b(x)$, $\g(x)$ and $P\sb L (x)$ such that
$$ \li{J^+ (x) &= \b (x)\cr
J^0 (x) &=-:\b(x)\g(x): - k_1 P\sb L (x)\cr
J^- (x) &= :\b(\s)\g^2(x): + 2k_1 \g(x) P\sb L (x) +
k_2 \g^{\prime} (x) \quad,&(4.11)\cr}$$
where
$$ [\b (x) ,\g (y) ] = -i \d (x-y) \quad,\quad
[ P\sb L (x) , P\sb L (y)] = i\d^{\prime} (x-y)
\quad,\eqno(4.12)$$
whith the other commutators bieng zero. Then the expressions (4.11)
satisfy the $SL(2,{\bf R})$ algebra (4.1) if
$$ k_1 = \sqrt{k +2\over 2}\quad,\quad k_2 = - k \quad,\eqno(4.13)$$
where the normal ordering in (4.11) is with respect to the Fourier modes
of $\b$ and $\g$. The scalar constraint now becomes
$$ \hat{{\cal S}} = :\b^{\prime}\g:  - \ha :P^2_L : +
{Q_L\over\sqrt{2\p}} P^{\prime}_L
 + \ha :P^2_D :+ {Q_D\over\sqrt{2\p}} P^{\prime}_D
+ \ha :P_i^2 : = 0 \quad,\eqno(4.14)$$
where
$$Q_L = k_1 - {1\over2k_1}\quad.\eqno(4.15)$$
Note that the transformation (4.11)
is also defined classically, with $k_1 ={\bar{\a}\over\sqrt2}$ and
$k_2 = -\bar{\a}^2$, satisfying the algebra (3.4).

If we define $B(x) = \b(x) - \l$ and $\G (x) = \g(x)$, then the $J^+$
constraint implies that $B =0$, and consequently we can drop the canonical
pair $(B,\G)$ from the theory. Therefore we are left with $P_L , P_D$ and
$P_i$ variables, obeying only one constraint
$$ {\cal S} = -\ha P_L^2 + {Q\sb L\over\sqrt{2\p}} P_L^{\prime}
 + \ha P^2_D + {Q_D\over\sqrt{2\p}} P^{\prime}_D + \ha P_i^2 =0
\quad.\eqno(4.16)$$

\sect{5. BRST Cohomology}

Now one needs to study the BRST cohomology of a Virasoro algebra
$$ L_n = \ha\su_m :\a_{n-m}\cdot\a_{m}: + inQ\cdot\a_n \quad,\eqno(5.1)$$
where
$$ X^{a}= (X^L , X^I )\quad,\quad X\cdot Y = \h_{ab}X^a Y^b \quad,
\quad \h_{ab} = {\rm diag}(-1,1,...,1) \quad,$$
$$ Q_a = (Q_L , Q_D, 0,...,0) \quad.\eqno(5.2)$$
The BRST charge is then given by the usual expression
$$ \hat{Q} = \su_n c_{n} L_{-n} + \ha \su_{m,n} (m-n):c_m c_n b_{-m-n}:
-c_0 a \quad.\eqno(5.3)$$
The normal ordering is with respect to the vacuum $\sket{vac}=\sket{p}\otimes
\sket{0}$
$$ \a_n\sket{vac}= c_n \sket{vac} = b_{n}\sket{vac} = 0 \quad,\quad n \ge 1
\quad,\eqno(5.4)$$
where $\sket{p}$ is the $\a$-modes vacuum ($\a_0\sket{p}=
p\sket{p}$), while $\sket{0}$ is the ghost vacuum, satisfying $b_0\sket{0}=0$
(the other possibility $c_0\sket{0}=0$ gives symmetric results).
Nilpotency of $\hat{Q}$ implies
$$ Q^2 = -Q^2_L + Q^2_D = 2 - N/12 \quad,\quad a = N/24 \quad,\eqno(5.5)$$
which are the conditions for the absence of anomalies. As expected, the first
condition in (5.5) is equivalent to (4.9) if $Q_L$ takes the value (4.15).

The zero ghost number cohomology is determined by the usual Gupta-Bleuler
conditions
$$ (L_n - a\d_{n,0})\sket{\j} = 0 \quad,\quad n\ge 0 \quad,\eqno(5.6)$$
where $\sket{\j}$ belongs to the $\a$-modes Fock space $F(\a)$.
Since the anomalies are absent, due to (5.5), then one can expect that the
classical counting of the physical degrees of freedom should hold. Hence
only the zero modes of gravity plus dilaton sector should survive, together
with N transverse local degrees of freedom, corresponding to the $P_i$ modes.
This can be verified for the first few levels.

Clearly, the ground state $\sket{p}$ is a solution of (5.6) if
$$p^2 =-p^2_L + p^2_I = N/12 \quad,\eqno(5.7)$$
which looks like the tachyionic ground state of the usual string theory.
At the first excited level, the states are of the form
$$ \sket{\j} = \x \cdot \a_{-1} \sket{p} \quad,\eqno(5.8)$$
which are physical if
$$ p^2 = N/12 - 2 \quad,\quad \bar{p}\cdot \x = 0 \quad,\eqno(5.9)$$
where $\bar{p} = p + iQ$. The norm of the state (5.8) is
$$ |\x|^2 = \x^*_a \x^a \quad,\eqno(5.10)$$
which for the physical states becomes
$$\li{ |\x|^2 =& S^{IJ}\x^*_I \x_J \cr
              =& \left( 1-{|\bar{p}_1|^2\over |\bar{p}_0|^2}\right) |\x_1|^2
 + \left( \d_{ij} - {p_i p_j\over |\bar{p}_0|^2} \right) \x^{*}_i \x_j
-\left( {p_i\bar{p}_1\over|\bar{p}_0|^2}\x^*_i \x_1 + c.c. \right)
\quad,&(5.11)\cr}$$
where $X_L =X_0$ and $X_D=X_1$.
By going into a special frame $p_0= m , p_1=p_i = 0$ for $N<24$ or
$p_0=p_1 = p , p_i =0$ for $N=24$ or $p_0 =p_1 =0, p_i^2 =m^2$ for $N>24$,
where $m^2 = |2-N/12|$, it is easy to see that the Hermitian
matrix $S^{IJ}$ always has $N$ positive eigenvalues and one zero eigenvalue.
Therefore this confirms our conjecture that only the transverse modes are
physical. This is different from the usual string theory ($Q^a=0$), where
$S^{IJ}$ is positive definite for $N=D-2 \le 24$, while for $N>24$, $S^{IJ}$
has negative eigenvalues, corresponding to the negative norm states.

Note that the so called ``discrete" states \cite{{liz},{bmp}}, arise when
$$\x_+ = -\x_- \quad,\quad \x_i = 0 \quad,\eqno(5.12)$$
where $\x_{\pm} = {1\over\sqrt2}(\x_0 \pm \x_1 )$, so that
$$ |\x|^2 = 2|\x_+|^2 >0 \quad.\eqno(5.13)$$
The physical state condition then implies
$$ \x_+ (\bar{p}_+ - \bar{p}_- ) = 0 \,\,\to\,\, \bar{p}_+ = \bar{p}_-
\,\,\to\,\,
p_{\pm} = -iQ_{\pm}\,\,\, {\rm or}\,\,\, p_{\pm} = i Q_{\mp}\,\,.\eqno(5.14)$$
However, the meaning of these states is not clear, since they have imaginary,
but fixed, momenta. See \cite{mik} for a more detailed discussion.

\sect{6. Conclusions}

We have solved non-perturbatively a class of 2d dilaton gravity theories,
defined by a potential $V(\f) = \L e^{\b\f}$. Note that the $\b\ne 0$ case is
essentially the same as the $\b= 0$ case analysed in \cite{mik},
since a set of
non-canonical variables, forming an $SL(2,{\bf R})\otimes U(1)$ algebra,
was found. In terms of these variables the constraints look the same in both
cases. The $SL(2,{\bf R})$ variables are gauge independent generalization
of the KPZ currents \cite{kpz}, which were originally found in the chiral
gauge for $\b=0$ theory. The Wakimoto construction (4.11) and the subsequent
free-field expression for ${\cal S}$ (4.16) is the canonical quantization
analog of the DDK construction \cite{ddk}.

We have given only a partial analysis of the BRST cohomology for the zero
ghost number sector. This analysis supports
our conjecture about the physical Hilbert space, which was based on the
classical counting of the physical degrees of freedom and the absence of
the anomalies.
We have also concluded that the physical Hilbert space is well defined
for any $N$, which conflicts with the conclussions of \cite{{strom},{alw}}.
There are several possible explanations for this discrepancy.
First, we have not done the full cohomology analysis. Second, their results are
semi-classical, and third, the theories may not be the same, since they
are working in the path-integral quantization scheme.

The kee question now is whether this exactly solvable model contains the
phenomena of interest, i.e. formation and evaporation of 2d black holes.
The 2d black hole classical solutions appear in the non-compact case, however
we will argue now that one can have black hole solutions even in the
compact case. As Mann et al.
have shown \cite{mann}, the original black hole solutions
correspond to a one sided colaps of a 1d dust, so that the singularity is
at $x=\pm\infty$. However, a symetric collaps produces a black hole solution
symmetric with respect to the origin, with the singularity at the origin $x=0$,
and the horizon at $x=\pm x_h$, $x_h <\infty$ \cite{mann}. By restricting
this solution to an interval
$[-L,L]$, such that $L>x_h$, we will obtain a black hole solution on a
compact interval.

A more sirious objection to our solvable model
as a model of quantum black holes is the
fact that we have used the Fourier modes of our fields
to define the quantum theory. A function can be expanded into a Fourier series
only if it is piecewise continious on $[-L,L]$. However, the black hole
metric blows up at the horizon, and therefore cannot be expanded into
a Fourier series. Hence by using the Fourier modes we are restricting the
phase space of our model to the space of piecewise continious solutions,
which does not contain the black hole solutions.
Strictly speaking,
this means that our model does not describe the phenomenon
of interest. However, one can hope that the Fourier modes construction
can be viewed as a some kind of a discrete approximation to the full theory.
Clearly, a further study is neccessary.

Therefore
one should find a way of defining the quantum theory without using the
Fourier modes. In the canonical quantization approach, one way would be to
recast the scalar constraint ${\cal S}$ into
a Schrodinger type equation. This would require defining an extrinsic time
variable, in analogy with the 4d quantum gravity \cite{kuch}.

If any of the proposed methods works, then a physical Hilbert space can be
constructed, and
therefore the quantum mechanics would stabilize a 2d black hole, in analogy
with the Hydrogen atom, which is classically unstable.

\end{document}